\documentclass[12pt, english]{article}

\usepackage[utf8]{inputenc}  
\usepackage[english]{babel}  
\usepackage[letterpaper, margin=1in]{geometry}  

\usepackage{natbib}
\usepackage[colorlinks,citecolor=black,urlcolor=blue,filecolor=blue]{hyperref}

\usepackage{bm}\usepackage{bbm}  
\usepackage{amssymb}  
\usepackage{amsmath}  
\usepackage{cases}     
\usepackage{multirow}  
\usepackage{longtable} 
\usepackage{caption}   
\usepackage{float}  

\usepackage{graphicx}  
\usepackage{setspace}  
\usepackage{array}     
\usepackage{xcolor}    
\usepackage[ruled,vlined]{algorithm2e}
\usepackage{url}       
\usepackage{adjustbox} 
\usepackage[title]{appendix} 

\begin{document}

\title{Clustering Structure of Microstructure Measures \thanks{We thank Prof. Gideon Saar, Dr. Manny Dong, the editors, referees and all the support from Cornell University.}
}

\author{
Liao Zhu\thanks{Department of Statistics and Data Science, Cornell University, Ithaca, New York 14853, USA.},
Ningning Sun\thanks{Department of Computer Science, Cornell University, Ithaca, New York 14853, USA.},
Martin T. Wells\thanks{Department of Statistics and Data Science, Cornell University, Ithaca, New York 14853, USA.}
}

\date{}
\maketitle
\global\long\def\thefootnote{\arabic{footnote}}


\begin{abstract}
This paper builds the clustering model of measures of market microstructure features which are popular in predicting the stock returns. In a 10-second time frequency, we study the clustering structure of different measures to find out the best ones for predicting. In this way, we can predict more accurately with a limited number of predictors, which removes the noise and makes the model more interpretable.
\end{abstract}

\section{Introduction}
Finding the features to explain and predict future stock returns has always been intensely studied issue in financial economic literature. For example, \cite{jegadeesh1993returns} show that a stock's past returns predict its future returns, \cite{jegadeesh1995short} showed that short-horizon return reversals are related to the liquidity, which was measured by the bid-ask spread. \cite{glosten1993relation} showed that the expected value and the volatility of the nominal excess return on stocks are related. \cite{chan2000trade} mentioned the role of the imbalance of trades and \cite{bessembinder1993price} discussed the relationship among price volatility, trading volume, market depth and stock returns. Apart from that, more market microstructure features can be potentially related, such as the Herfindahl-Hirschman index in \cite{herfindahl1950concentration}.

Furthermore, there are always different measures created for each of these features. Take the liquidity as an example, the most commonly used measure is the bid-ask spread. However, the most commonly used measure may not always be the best one for all purpose. Some other versions of measure of liquidity may be more insightful, such as the the effective bid-ask spread, see \cite{roll1984simple}. Even for the same measure, taking the last-prevailing value and the time-weighted value may still be different.

With the increase of trading frequency, volume, and the boom of technology, information are more likely to be captured faster. In the paper by \cite{chinco2017sparse}, a simple LASSO model was applied to choose significant factors from about 6000 stocks' 1-minute ahead returns as the sparse and short-lived signals. This suggests the need to use short-term measures instead of long-term ones to predict short-term future returns. Instead of using only the returns from stocks as candidate features to predict short-term future returns, there are many other features that can be taken into account, as stated above. For example, in a 10-second horizon, the liquidity may play an important role in predicting the future returns.

Considering the above, instead of the traditional way of studying the impact of one feature on long-term future stock returns, this paper investigates the effect of a larger number of them together in a short horizon. Considering that there are a number of features related, for each feature there are several different measures, and furthermore, several kinds of ways to calculate each measure, we are in fact facing a large number of candidate measures to choose from. Since some of these measures are related to the same feature, and some features may be strongly related, the measures are sometime highly correlated and will cause dimension issue and redundancy if we use them all. For further prediction of one stock return, we need to consider the impact from a large number of different stocks, which will be in a high-dimensional regime. Any redundant measures included for each company can cause large difficulty for further analysis.

Considering these difficulties, it is worthwhile to use the high-dimensional statistical methods to reduce the dimension and find out the good representatives of the measures. In the meantime, by high-dimensional statistical methods, the relationship between different features can also be studied. The most traditional method for dimension reduction is the Principal Component Analysis (PCA) method. However, in our case, the covariance decomposition type of methods (including the PCA) has some intrinsic weakness since it can mix the underlying features together or separate them into several principal components. In addition, PCA may ignore some features that really matters since this type of methods are based on the variance. A feature with low variance and low loadings in the principal components (and thus ignored by the PCA) can still have a high correlation with the future stock returns. This issue can be well solved by the prototype-clustering since the prototypes are selected based on the correlation, not variance. Using the prototype clustering method to form the clustering structure of the measure, we can have a clear insight and interpretation of the relationship among different measures. The similar methods were also used in fitting the Adaptive Multi-factor Models, see \cite{zhu2020high, zhu2021time, jarrow2021low}.

In this paper, we first list all the calculation of the the measures related predicting in Section \ref{sec_measures}. Then a brief discussion of the statistical methods used is included in Section \ref{sec_stats}. Section \ref{sec_analysis} describes our analysis procedure and results and Section \ref{sec_conclusion} concludes.

\section{Related Features and Measures}
\label{sec_measures}
Here is the list of all features and the measures we use to describe them.

\subsection{Returns}
The first is the return. We use mid-quote prices instead of trade prices since trade prices are only effective at the time spot when it occurs and are not effective before nor after. However, quotes prices are almost continuous since they are effective for a time period before a trade ends it. Therefore the equation we use is
\begin{equation}
r_t = \frac{q_t}{q_{t - \Delta t}} - 1
\end{equation}
where $r_t$ means the return and $q_t$ means the mid-quote at time t. In this paper we use frequency 10 seconds as our $\Delta t$. The mid-quotes are the last prevailing values.

\subsection{Bid-ask spreads}
For the liquidity, we include different measures of the spreads. For each measure, we consider both prevailing and time-weighted values. For each 10-second time interval, the measures for spreads are:
\begin{enumerate}
  \item Dollar bid-ask (quoted) spread $ = \text{ask price} - \text{bid price} $
  \item Proportional bid-ask (quoted) spread $ = \frac{\text{ask price} - \text{bid price}}{\text{mid quote}} \times 100\%$
  \item Dollar effective spread:  $ 2|\text{trade price} - \text{mid quote}|$.
  \item Proportional effective spread: $ 2\Big|\frac{\text{trade price} - \text{mid quote}}{\text{mid quote}}\Big| \times 100\%$.
\end{enumerate}

\subsection{Volatility of prices}
For volatilities, since we are in a short horizon, the standard deviation of returns does not seem reasonable since there may not be enough records for accurate estimation. So we use the (high - low) for bid prices, ask prices, mid-quote, or trade prices within the 10-second time interval, normalized by the Average Daily Realized Volatility (ADRV) defined by
\begin{equation}
  \text{ADRV} = \text{daily ask high} - \text{daily bid low}
\end{equation}
of the prevailing month as measures of volatility. The daily data are from the CRSP database.

\subsection{Measures related to trades}

To have a measure related to the imbalance of trades, we should first classify the trades into buyer-initiated trades and seller-initiated trades. The paper \cite{chakrabarty2006trade} gives a nice brief review of these methods and proposed their own improved algorithm. The popular algorithms are: LR algorithm by \cite{lee1991inferring}, EMO algorithm by \cite{ellis2000accuracy}, \cite{chakrabarty2006trade}. Here we use the CLNV algorithm \cite{chakrabarty2006trade} since it is slightly more accurate.

The measures of trade frequency includes the count number of the all trades within each 10-second time period, and the average time between trade.

The measures of the trade volume are the last prevailing / time-weighted average value of: dollar amount / number of shares / number of shares normalized by ADRV of the prevailing month.

For the imbalance of trades, for each 10-second time period, the equations are: \\
\begin{equation}
  \frac{\text{buyer - seller}}{\text{buyer + seller}} \text{   (directional)   or   }
  \Big|\frac{\text{buyer - seller}}{\text{buyer + seller}}\Big|   \text{  (nondirectional)  }.
\end{equation}
And we can use the previous measures for trade frequencies and trade volumes for the buyer and seller in the equations above.

\subsection{Measures related to quotes}

The measures of the quote frequencies can be the count number of: all quote records / bids changes / asks changes, and the average time between them quote records/ bids changes/ ask changes within the 10-second intervals. The ``quote change'' measures are measures that only count for quotes with different quote prices, as a measure of new information. Note that for the count number of bid changes, we will count the consecutive same bid prices as only 1 count. Similarly for asks. We also include average time between quotes / quote changes here. It is almost reciprocal (multiplied by a constant) of the count number. But since it is not a linear function of count number, it is totally different from the count number in regression.

The depths of the market are based on the quotes for each stock, as a measure of liquidity and an indicator of price movement direction. The measures related to the depth of the market can be:
\begin{enumerate}
  \item The last prevailing ask / bid / ask - bid / $|\text{ask - bid}|$.
  \item The time weighted values, specifically,\\
      $ \int{\text{ask}_t dt} $, $ \int{\text{bid}_t dt} $, $ \int{(\text{ask}_t - \text{bid}_t) dt} $, $  \int{|\text{ask}_t - \text{bid}_t| dt} $.
\end{enumerate}
For each ask (or bid) in the equation above, we can use dollar volume / number of shares / number of shares normalized by Average Daily Trading Volume (ADTV) of the prevailing month. And we can consider the measure for quotes in each exchange or the best quote nation-wide.

We also include the imbalance of quotes using a similar expression with that of imbalance of trades. For each 10-second time interval, the equations are: \\
\begin{equation}
  \frac{\text{ask - bid}}{\text{ask + bid}} \quad \text{   (directional) \quad  or \quad  }
  \Big|\frac{\text{ask - bid}}{\text{ask + bid}}\Big|   \quad \text{  (nondirectional)  }.
\end{equation}
And we can use the previous measures for quote frequencies and depth for the Ask and Bid in the equations above. In addition to calculating the time weighted value and then plug in the fraction as we did before, we shall also compute the time-weighted averages of the fractions:\\
\begin{equation}
  \int \frac{\text{(ask size)}_t - \text{(bid size)}_t}{\text{(ask size)}_t + \text{(bid size)}_t} dt \quad \text{ and } \quad
  \int \Big| \frac{\text{(ask size)}_t - \text{(bid size)}_t}{\text{(ask size)}_t + \text{(bid size)}_t} \Big| dt.
\end{equation}
Note that since we have them in both nominator and denominator, normalizing by ADTV will not make any difference now, in other words, these measures are already normalized.

\subsection{Concentration among exchange places}
The concentration among exchange places can be measured by the Herfindahl-index (HHI) of concentration, which is defined as:
\begin{equation}
    \text{HHI}_v=\sum_{i=1}^N v_i^2
\end{equation}
where $v_i$ means the fraction of value of the i-th exchange.

The original measure didn't use the fraction, but directly use the shares of each part. However, in our case, using fraction should be better because we don't want the difference of shares, volume, etc across different stocks been taken into account here, since they are taken care of in other measures. And since we use the HHI to measure the concentration of trades and quotes for each single company, normalizing by average daily value is no longer needed.  For each 10-second time interval, the choices of values ($ v_i $) can be the measures of trade frequency, trade volume, quote frequency, or the depth.

Considering all the measures above, we have more than 100 measures in total, and lots of them are very correlated. Therefore, we use the statistical methods to remove redundancies and find out good representatives.

\section{Statistical Methods}
\label{sec_stats}

This section describes the prototype clustering to be used to efficiently deal with the problem of high correlation among the measure. To remove unnecessary independent variables, using clustering methods, we classify them into similar groups and then choose representatives from each group with small pairwise correlations.
First, we define a distance metric to measure the similarity between points (in our case, the returns of the independent variables). Here, the distance metric is related to the correlation of the two points, i.e.

\begin{equation}
d(r_{1},r_{2})=1-|corr(r_{1},r_{2})|\label{CorrDist}
\end{equation}
where $r_{i}=(r_{i,t},r_{i,t+1},...,r_{i,T})'$ is the time series
vector for independent variable $i=1,2$ and $corr(r_{1},r_{2})$
is their correlation. Second, the distance between two clusters needs
to be defined. Once a cluster distance is defined, hierarchical clustering
methods (see \cite{kaufman2009finding}) can be used to organize the
data into trees.

In these trees, each leaf corresponds to one of the original data
points. Agglomerative hierarchical clustering algorithms build trees
in a bottom-up approach, initializing each cluster as a single point,
then merging the two closest clusters at each successive stage. This
merging is repeated until only one cluster remains. Traditionally,
the distance between two clusters is defined as either a complete
distance, single distance, average distance, or centroid distance.
However, all of these approaches suffer from interpretation difficulties
and inversions (which means parent nodes can sometimes have a lower
distance than their children). To avoid these difficulties, \cite{bien2011hierarchical} introduced hierarchical clustering with prototypes via a minimax linkage
measure, defined as follows. For any point $x$ and cluster $C$, let
\begin{equation}
d_{max}(x,C)=\max_{x'\in C}d(x,x')
\end{equation}
be the distance to the farthest point in $C$ to $x$. Define the
\emph{minimax radius} of the cluster $C$ as
\begin{equation}
r(C)=\min_{x\in C}d_{max}(x,C)
\end{equation}
that is, this measures the distance from the farthest point $x\in C$
which is as close as possible to all the other elements in C. We call
the minimizing point the \emph{prototype} for $C$. Intuitively, it
is the point at the center of this cluster. The \emph{minimax linkage}
between two clusters $G$ and $H$ is then defined as
\begin{equation}
d(G,H)=r(G\cup H).
\end{equation}
Using this approach, we can easily find a good representative for
each cluster, which is the prototype defined above. It is important
to note that minimax linkage trees do not have inversions. Also, in
our application as described below, to guarantee interpretable and
tractability, using a single representative independent variable is
better than using other approaches (for example, principal components
analysis (PCA)) which employ linear combinations of the independent
variables.

Similar approaches can also be found in the literature related to the Adaptive Multi-factor Model, see \cite{zhu2020high, zhu2021news, zhu2021time, jarrow2021low, zhu2020adaptive}. This method can be potentially used with time series analysis, see  \cite{zhao2020online, zhao2021group}, and some other machine learning models, such as \cite{huang2021staying, huang2020time, li2021frequentnet, jie2018stochastic, jie2018decision, zhang2021form, du2020multiple, mao2020differentiate, mao2020speculative, stein2021quclassi, stein2020hybrid, stein2021qugan, du2021improved, li2021two, davis2021clustering, wang2020efficient, kong2018study, bo2021subspace}.

\section{Analysis Procedure and Results}
\label{sec_analysis}

We use the Daily TAQ data base to form all the measures for each 10-second period. The average daily values (average daily trading volume, average daily realized volatility, etc.) are calculated for the prevailing month from the CRSP database. In this paper we use the day Apr. 3rd, 2018 as an example, and all the daily average value is calculated in the prevailing month, which is from Mar. 2nd, 2018 to Apr. 2nd, 2018.

\begin{figure}
\vspace{-1in}
\hspace*{0in}
\includegraphics[height = 0.9\textwidth, angle = 270]{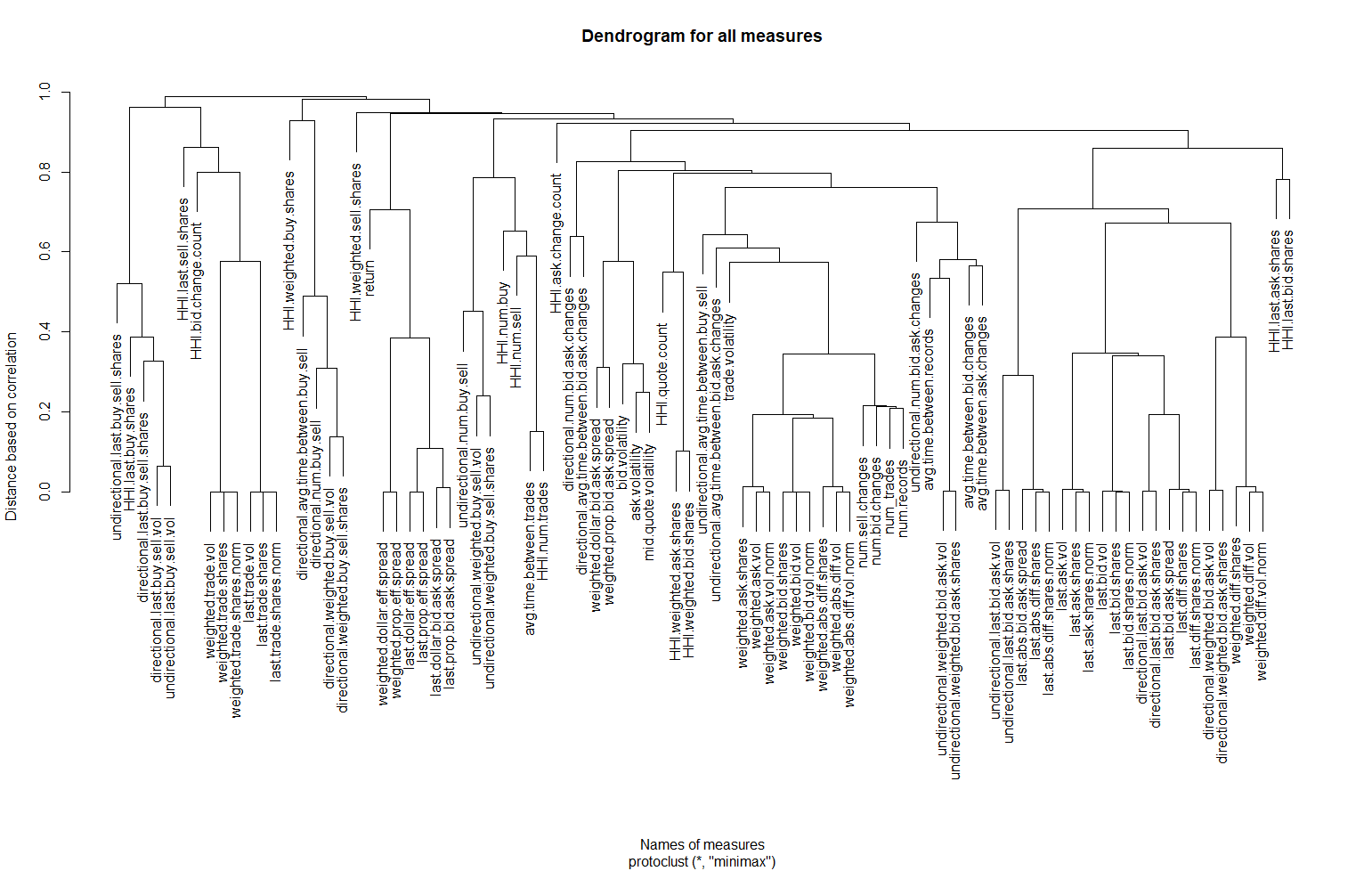}
\vspace{-0.5in}
\caption{Dendrogram of all measures.}
\label{fig1}
\end{figure}

We calculate the distance between measures by the equation \ref{CorrDist} for each of company among all 7263 companies in the database. Then we take the average of distances of each pair of measures over all the companies. The first prototype clustering dendrogram result is in Figure \ref{fig1}.  There are 91 measures in total, but some of them are nearly perfectly correlated. The following are the reasons for the strong correlation between some measures.

From the dendrogram, it is clear that the measures related to dollar volume always has a strong correlation with that related to the number of shares, since the price within a day does not change very much. Since we already have measures related to prices, here we should always use the measures related to the number of shares and remove redundant ones based on the dollar volume.

Also, the dollar effective spread always has strong correlation with proportional effective spread, since mid-quotes does not change much during one day. Therefore, in short horizon, we can abandon the dollar effective spread and only use the proportional effective spreads since they have normalization and will be comparable across the stocks.

\begin{figure}
\vspace{-1in}
\hspace*{0in}
\includegraphics[height = \textwidth, angle = 270]{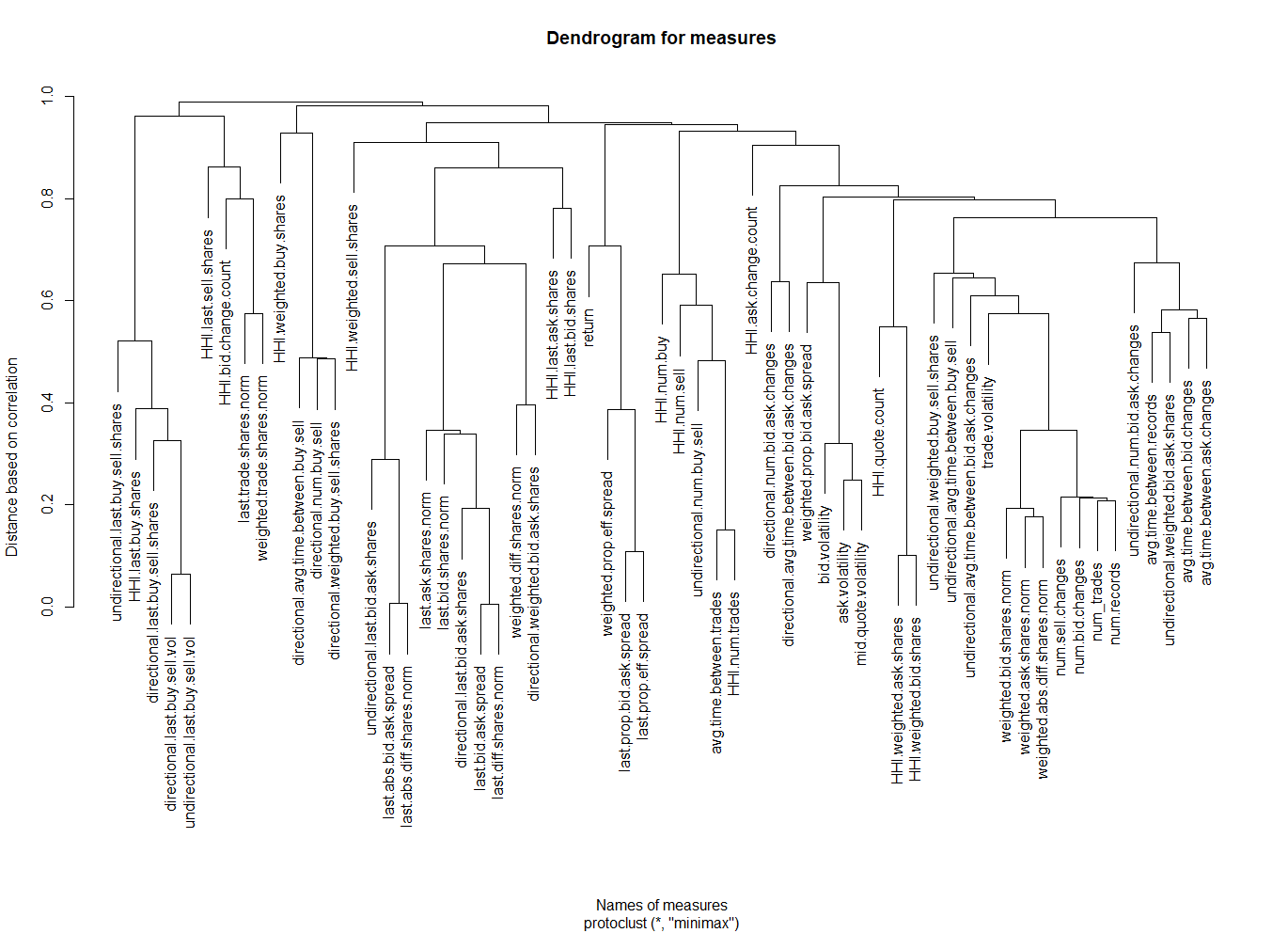}
\vspace{-0.5in}
\caption{Dendrogram of measures after removing redundant ones.}
\label{fig2}
\end{figure}

Apart from the normalizing issue above, the measure normalizing by the average daily trading volume (or daily volatility) has correlation 1 with the original measure. So we only need one of each normalized / non-normalized pair. Since we want to make measures stable across all stocks, it is better to use the normalized ones. Therefore we should remove all non-normalized measures if there is an appropriate normalized one.  The dendrogram after removing these redundant measures looks neater in Figure \ref{fig2}.

There are no longer perfect correlated measures. However, there are still some interesting discoveries that is worthy for further study. For example, the last prevailing bid-ask spread (which is the last prevailing value of (ask price - bid price)) is strongly correlated with the
last prevailing (ask shares - bid shares) / average daily trading shares. This suggests the relationship between quote prices and quote shares.

\begin{figure}
\vspace{-1in}
\hspace*{0in}
\includegraphics[width = \textwidth, angle = 0]{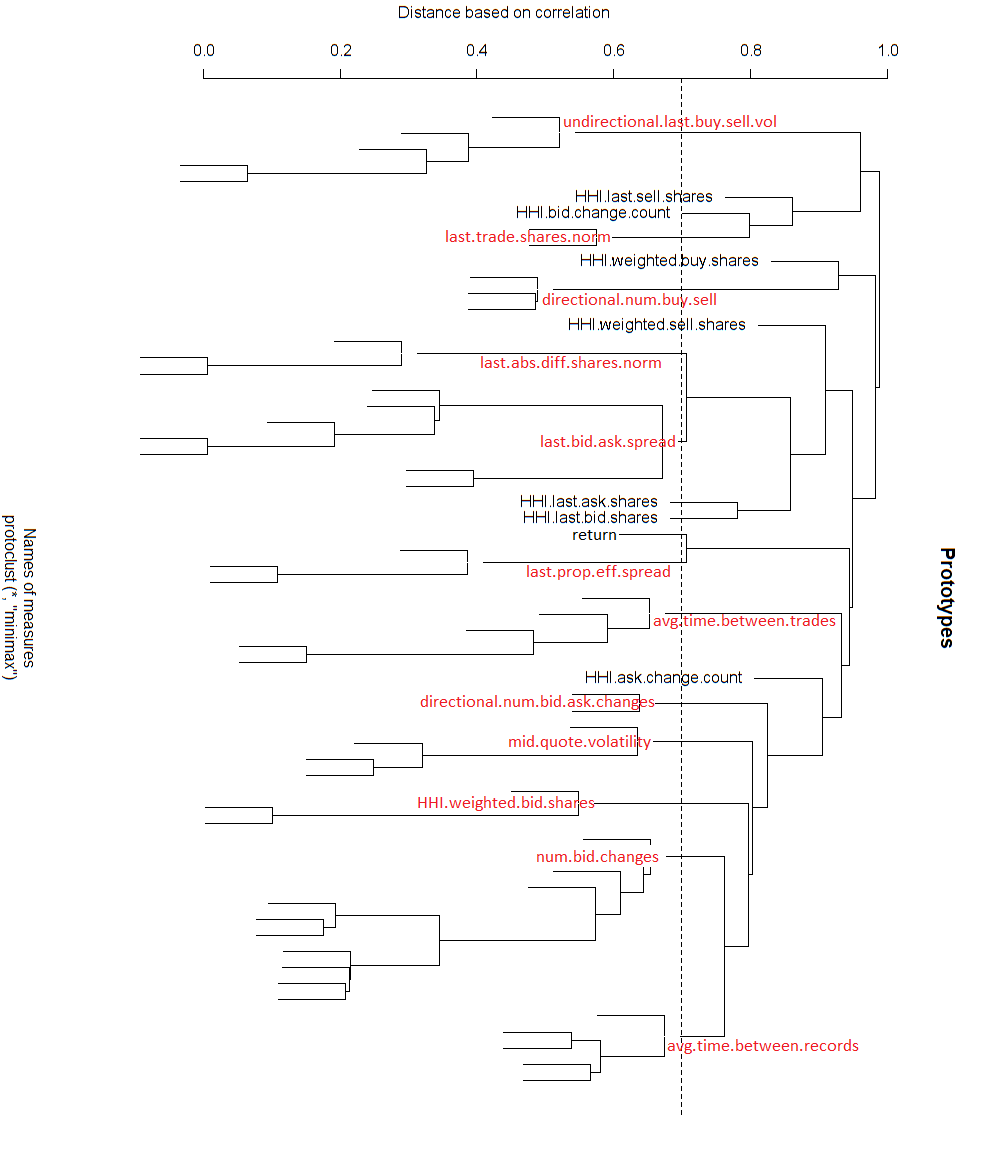}
\vspace{-0in}
\caption{Prototypes selected.}
\label{fig3}
\end{figure}

\begin{table}
\center %
\caption{List of all selected measures.}
\begin{tabular}{lll}
\hline
No.	&	Names	                   &	Descriptions	\\\hline
1	&	return	                   &	Return	\\\hline
2	&	last.prop.eff.spread	   &	The last prevailing proportional effective spread	\\\hline
3	&	mid.quote.volatility	   &	Volatility based on mid-quote	\\\hline
4	&	num.bid.changes	           &	The number of bid changes	\\\hline
5	&	avg.time.between.trades	   &	Theaverage time between trades 	\\\hline
6	&	last.trade.shares.norm	   &	Thelast prevailing trade shares normalized by	\\
	&		                       &	The average daily trading shares	\\\hline
7	&	directional.num.buy.sell   &	(buy - sell) / (buy + sell)  where buy (sell) in 	\\
	&		                       &	the equation means the number of buy (sell)	\\
    &                              &    trades \\\hline
8	&	undirectional.last.buy.sell.vol	
                                   & $|$buy - sell$|$ / $|$buy + sell$|$ where buy (sell) in \\
	&		                       & the equation means shares of buy (sell) trades	\\\hline
9	&	avg.time.between.records   &	The average time between quote records	\\\hline
10	&	last.bid.ask.spread	       &	The last prevailing bid-ask spread	\\\hline
11	&	last.abs.diff.shares.norm  &	The last prevailing (ask shares - bid shares) \\
	&		                       &    normalized by the Average daily	trading\\
    &                              &    shares over the last month \\\hline
12	&	directional.num.bid.ask.changes	
                                   &	(ask - bid) / (bid - ask) where ask (bid) in \\
	&		                       &    the equation means the number of ask (bid) \\
    &                              &    changes\\\hline
13	&	HHI.last.sell.shares	&	HHI of last sell prevailing shares	\\\hline
14	&	HHI.weighted.buy.shares	&	HHI of time-weighted buy shares	\\\hline
15	&	HHI.weighted.sell.shares	&	HHI of time-weighted sell shares	\\\hline
16	&	HHI.last.ask.shares	&	HHI of time-last ask shares	\\\hline
17	&	HHI.last.bid.shares	&	HHI of last prevailing bid shares	\\\hline
18	&	HHI.weighted.bid.shares	&	HHI of time-weighted bid shares 	\\\hline
19	&	HHI.bid.change.count	&	HHI of count of bid changes 	\\\hline
20	&	HHI.ask.change.count	&	HHI of count of ask changes	\\\hline
\end{tabular}
\label{tab1}
\end{table}

Finally, we use the prototype clustering to find good representatives (prototypes) of measures within each cluster, removing all redundant measures with correlation more than 0.3, in other word, distance less than 0.7).  The dendrogram of the prototypes selected is shown in Figure \ref{fig3}. There are 20 measures (out of 91 original measures) selected. The list of selected measures is in Table \ref{tab1}.

\section{Conclusion}
\label{sec_conclusion}

Using the high-dimensional statistical methods, we find the 20 representative measures from all 91 candidates and meanwhile discover the relationship between measures. A clear clustering structure is given by the dendrogram. The clustering structure gives a clear interpretation of the representatives and relationship between different measures. Also, interesting phenomena can be found and can be studied within the clustering structure. The clustering structure provides the flexibility of modifying the number of features to use in future studies.

Future work can be done to use the selected measures to fit more exotic statistical models, such as time series models, machine learning models etc., to predict the stock returns.

\bibliographystyle{apa}   
\bibliography{bib_liao}

\end{document}